# A nonlinear interface for phase transduction

LINSHAN SUN[1,*], HAO ZHANG[1,2], CAMERON LEARY[1], ALEX AMADOR[1], AND SERGIO CARBAJO[1,2,3,4,*]

[1]*Department of Electrical & Computer Engineering, University of California Los Angeles, Los Angeles, CA 90095, USA*
[2]*SLAC National Accelerator Laboratory, Stanford University, Menlo Park, California 94025, USA*
[3]*California NanoSystems Institute, 570 Westwood Plaza, Los Angeles, CA 90095, USA*
[4]*Physics and Astronomy Department, University of California, Los Angeles, CA 90095, USA*
*lssun@ucla.edu
*scarbajo@ucla.edu

**Abstract:**

Broadband nonlinear frequency conversion mixes multiple input frequencies, complicating the transfer of programmed spectral phase between widely separated spectral bands. Here we identify and experimentally realize a quasi-linear phase-transduction regime enabled by broad phase-matching acceptance and effective localization of the complex frequency-mixing response. This selectivity can arise from coherent interference without narrowing the pump spectrum. We implement this regime through dispersive four-wave mixing in an argon-filled hollow-capillary fiber, transferring spectral phase from 1032-nm near-infrared signals to 344-nm ultraviolet idlers. Transient-grating frequency-resolved optical gating reveals frequency-inverted, phase-conjugating transfer of quadratic dispersion, localized phase steps, and low-frequency sinusoidal modulation. After correction for the modeled pump-induced phase background, the quadratic and step responses exhibit approximately linear point-by-point correlations. With increasing pump chirp, the transferred phase steps become more sharply preserved, and the fitted phase-transfer gains approach unity, consistent with enhanced effective localization. These results demonstrate quasi-linear spectral-phase transduction in a co-propagating broadband nonlinear interaction without spatially separating its spectral components.

Spectral phase is a central degree of freedom of an ultrafast optical field[1]. For a given spectral intensity, it determines interference among frequency components and thus the temporal waveform, instantaneous frequency, and coherent evolution[2–6]. Direct phase-shaping technologies span the wavelength from the infrared to the ultraviolet (UV), but their spectral resolution, usable bandwidth, optical throughput, and pulse-energy handling remain strongly wavelength dependent[7–10]. These limitations are particularly restrictive for broadband, high-energy UV pulses, for which direct phase control is often more difficult than pulse generation[11–19]. A phase-preserving interface between spectral regions would allow a waveform to be programmed where phase control is most effective and transferred to the target wavelength[17,20–23].

Nonlinear frequency conversion is a natural candidate for such an interface, but coherence alone does not guarantee direct phase transduction[17,21,22,24–27]. In a broadband interaction, each output frequency generally receives coherent contributions from multiple combinations of input frequencies. Their interference can smooth localized phase features, redistribute phase modulation into spectral amplitude, or add a process-dependent phase background[24,25]. Previous experiments have transferred chirp and programmed phase masks through sum-frequency generation and four-wave mixing, including indirect shaping at UV wavelengths[17,21,24,25,28]. Frequency-domain nonlinear optics instead physically separate spectral components before conversion, allowing each to undergo an approximately independent, quasi-monochromatic interaction[24,25,29]. This suppresses the spectral convolution intrinsic to conventional broadband mixing and can enable nearly linear phase transfer[24,25]. Whether broadband nonlinear frequency conversion can achieve near one-to-one spectral-phase mapping without spatially separating the interacting frequencies remains unclear.

Such a mapping can emerge when the phase-matching response remains broad across the participating bandwidths[30–32] and the pump-conditioned mixing kernel provides effective localization around a well-defined input-to-output frequency trajectory. While a spectrally narrow pump can achieve this selectivity by restricting the range of contributing input frequencies, spectral narrowing is not a necessary condition. A broadband pump can instead produce an effectively localized sampling response through coherent interference within the complex mixing kernel, without narrowing its spectral magnitude. In this limit, the broadband convolution approximately preserves the mapped input phase, up to a deterministic background and process-dependent phase conjugation[17,22,25].

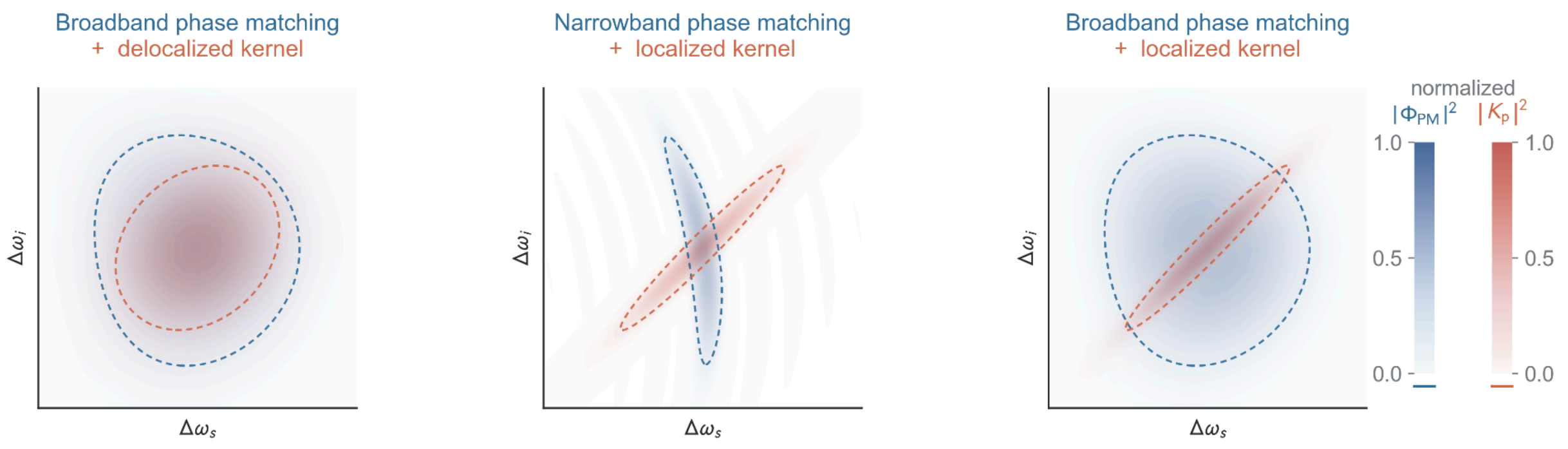

**Fig. 1 | Conditions for quasi-linear spectral-phase transduction in broadband nonlinear frequency conversion.** Conceptual representation of the phase-matching response and effective pump-conditioned frequency selectivity in the input–output frequency plane. The schematic effective pump kernel ($K_p$) represents the frequency selectivity of the mixing process, which may arise from spectral-amplitude confinement or phase-dependent coherent interference. **Left**, broad phase-matching acceptance ($\Phi_{PM}$) alone is insufficient when the effective mixing response is delocalized, because each output frequency receives contributions from a broad range of input frequencies. **Middle**, a localized effective mixing response can provide frequency selectivity, but a narrow phase-matching response restricts the usable conversion bandwidth. **Right**, the combination of broad phase-matching acceptance and effective kernel localization around an input-to-output frequency trajectory enables broadband, approximately point-to-point spectral mapping. In this regime, the programmed input spectral phase can be retained at the output, apart from a deterministic background and any process-dependent phase conjugation. Dashed contours indicate the half-maximum boundaries of the normalized phase-matching response and schematic effective pump kernel.

Here we identify and experimentally realize quasi-linear spectral-phase transduction using dispersive four-wave mixing (DFWM) in an argon-filled hollow-capillary fiber (HCF)[12,16,33–36]. The interplay between pressure-dependent gas dispersion and capillary waveguide dispersion provides broadband phase-matching acceptance, as characterized previously in Ref. 17. A chirped 516-nm pump provides effective frequency selectivity for transferring programmed spectral phase from a 1032-nm signal to a 344-nm idler along the frequency-inverted trajectory $\omega_i \simeq 2\omega_p - \omega_s$. Using transient-grating frequency-resolved optical gating (TG-FROG), we reconstruct the input and output complex fields and demonstrate phase-conjugating transfer of quadratic dispersion, localized phase steps, and low-frequency sinusoidal modulation. After correction for the modeled pump-induced phase background, the quadratic and localized-step measurements exhibit approximately linear point-by-point phase correlations. Increasing pump chirp further improves the preservation of localized phase steps and brings the fitted phase-transfer gain closer to unity, consistent with enhanced effective localization of the mixing response. These results demonstrate that a co-propagating broadband nonlinear interaction can support quasi-linear spectral-phase transduction without requiring a spectrally narrow pump or spatial separation of frequency components.

## Quasi-linear spectral-phase transduction

For fixed pump fields, weak depletion, and negligible self-action of the shaped input, the dependence of a phase-coherent parametric process on the shaped input can be written as a linear or conjugate-linear field transformation,

$$\tilde{E}_{\text{out}}(\Omega) = \int H(\Omega, \omega)\tilde{E}_{\text{in}}^{(\sigma)}(\omega)\, d\omega \tag{1}$$

where $H(\Omega, \omega)$ contains the pump-conditioned mixing response, phase matching and propagation. $\tilde{E}_{in}^{(\sigma)}$ denotes either the input field or its complex conjugate, depending on the process. For phase transduction, the relevant distinction is not this formal field linearity itself, but whether the kernel is sufficiently localized around a one-to-one input-to-output frequency trajectory. When

$$H(\Omega, \omega) \simeq G(\Omega)\delta[\omega - \omega_m(\Omega)] \tag{2}$$

the broadband convolution reduces to an approximately point-to-point mapping, which is the ideal local-sampling limit,

$$\varphi_{\text{out}}(\Omega) \simeq \varphi_G(\Omega) + s\varphi_{\text{in}}[\omega_m(\Omega)] \tag{3}$$

Here, $s = \pm 1$ distinguishes direct and phase-conjugating transfer. For a chirped broadband pump, effective localization arises from coherent cancellation in the integral action of the complex mixing kernel, rather than from a narrowing of its spectral magnitude, as discussed in Supplementary Note 2.

### SOD mapping

We first benchmark the phase-transfer response using DFWM in an argon-filled HCF. As shown in **Fig. 2a**, a programmable 1032-nm signal is combined with a chirped 516-nm pump to generate a 344-nm idler. The HCF provides transverse confinement and a phase-matching acceptance spanning the interacting bandwidths, while the chirped pump provides the time–frequency selectivity[37] required for a localized signal-to-idler mapping. The underlying mechanism and its dependence on pump GDD are examined in Supplementary Note 2 and Fig. S2. The complex signal and idler fields are independently reconstructed by TG-FROG.

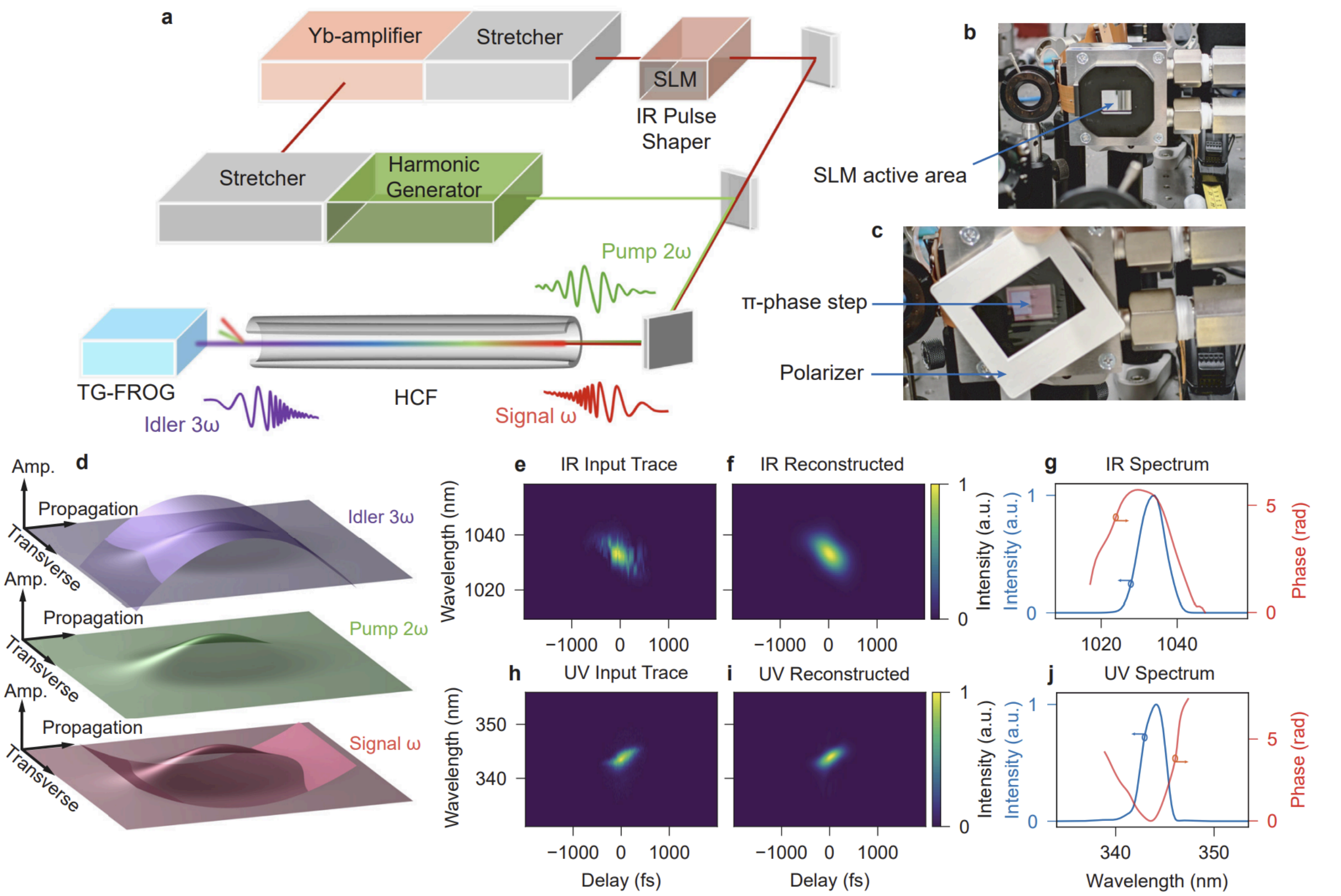

**Fig. 2. Experimental configuration and SOD phase mapping. a,** Schematic of the experimental setup. The signal (ω) is shaped by a programmable IR pulse shaper and combined with the pump (2ω) before entering the gas-filled hollow-core fiber (HCF). The generated idler (3ω) is characterized via TG-FROG. **b, c**, Photographs of the IR pulse-shaper setup. **b**, SLM with the water-cooling system. **c**, View through the polarizer with a central π-step phase applied to the SLM. **d**, Conceptual illustration of spectral-phase transfer during nonlinear propagation in the HCF, where the signal spectral phase is mapped to the generated idler through broadband four-wave mixing. **e, f**, Measured and reconstructed TG-FROG traces of the input signal with applied second-order dispersion (SOD). **g**, Retrieved spectral amplitude and phase of the signal. **h, i**, Measured and reconstructed TG-FROG traces of the generated idler. **j**, Retrieved spectral amplitude and phase of the idler.

We begin with a smooth quadratic spectral phase as a benchmark for the transfer response. A GDD of $-22845\ \mathrm{fs}^2$ is applied to the signal, while the pump carries approximately $3 \times 10^4\ \mathrm{fs}^2$. The measured and reconstructed TG-FROG traces and the retrieved complex spectra are shown in **Fig. 2e–j**. A calculation model for chirped DFWM predicts an idler GDD of $9294\ \mathrm{fs}^2$ (Supplementary Note 4B), compared with $6194\ \mathrm{fs}^2$ retrieved experimentally. The measured idler GDD is smaller than the value predicted by the reduced chirped-DFWM model. Possible contributions from nonlinear propagation and retrieval uncertainty are considered below. We next test whether the same frequency-inverted phase mapping persists for phase structures that cannot be represented by a single dispersion coefficient.

### Localized phase-step transfer

The smooth SOD case probes the low-spectral-modulation-frequency limit of the phase response. We next test the localized-kernel picture with a sharper feature by imposing a nominal π-step on one side of the NIR spectrum. The pump conditions are unchanged, so the signal carries the same smooth dispersive background together with a localized phase transition. Two additional step positions are examined in the SI.

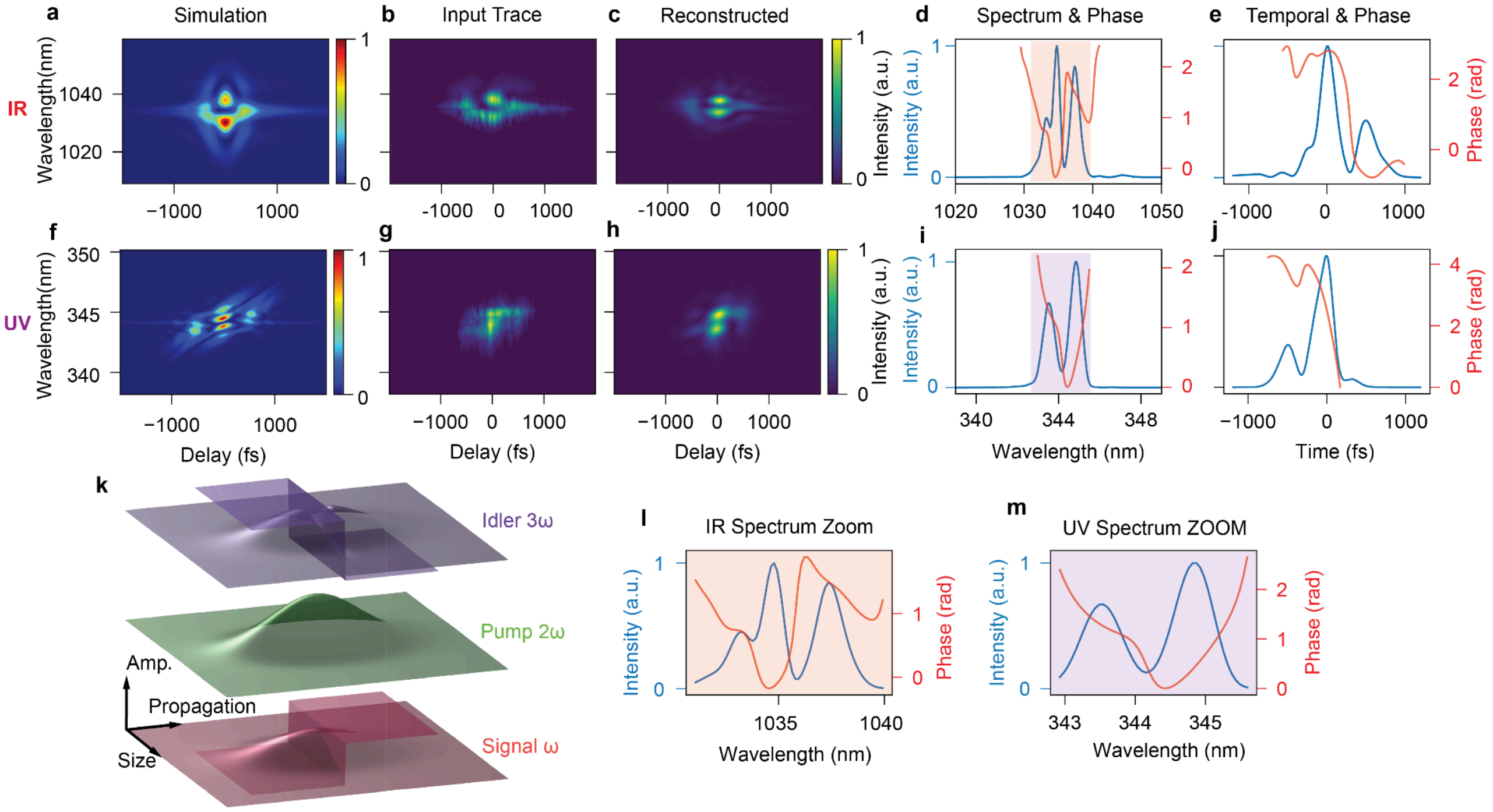

**Fig. 3 | Nontrivial spectral phase transfer with a π-step. a–e**, Simulated and experimental characterization of the near-infrared signal pulse carrying a π-step phase on top of SOD. **a**, Simulated TG-FROG trace. **b,c**, Measured and reconstructed TG-FROG traces. **d**, Retrieved spectral amplitude and phase. **e**, Time-domain intensity and instantaneous phase. **f–j**, Corresponding characterization of the generated ultraviolet idler pulse. **f**, Simulated TG-FROG trace. **g,h**, Measured and reconstructed TG-FROG traces. **i**, Retrieved spectral amplitude and phase. **j**, Time-domain intensity and instantaneous phase. **k**, Schematic illustration of the transfer process in the hollow-core fiber. The pump carries only SOD, whereas the signal carries both SOD and a localized π-step. The idler retains features of both contributions. **l,m**, Zoomed views of the spectral amplitude and phase in **d** and **i**, showing that the step feature is transferred from the long-wavelength side of the signal to the short-wavelength side of the idler, as expected for the $2\omega_p - \omega_s$ DFWM interaction.

The measured and reconstructed TG-FROG traces and retrieved spectral fields (**Fig. 3b–d,g–i**) show transfer of the step while retaining the overall dispersive background. In the zoomed spectra (**Fig. 3l,m**), the step on the longer-wavelength side of the signal appears on the shorter-wavelength side of the idler, consistent with the frequency-inverted $2\omega_p - \omega_s$ mapping. The programmed 3.1 rad step is retrieved as approximately 2.1 rad in the NIR signal and 1.1 rad in the UV idler. The UV transition is also broader, and the step appears on the opposite side of the spectrum, as expected from the frequency-inverted mixing relation. The reduced contrast may reflect both finite mixing bandwidth and smoothing by the UV diagnostic. These contributions are addressed further through the calibrated comparison below.

The corresponding temporal reconstructions are shown in **Fig. 3e,j**. A one-to-one temporal mapping is not expected because the transduction relation is defined spectrally and the generated idler amplitude remains pump conditioned.

A joint SOD-plus-step fit gives an idler GDD of approximately $7.7 \times 10^3\ \mathrm{fs}^2$ and a transferred step amplitude of approximately $1.1$ rad. The corresponding prediction for the GDD is $1.45 \times 10^4\ \mathrm{fs}^2$. Although the measured quadratic curvature is smaller than predicted by the reduced model, the idler simultaneously retains the smooth dispersive background and the frequency-inverted localized phase transition. The mapping is therefore not restricted to a single low-order dispersion parameter. Its quantitative phase-transfer response, including the attenuation and broadening of the localized feature, is examined below.

Using matched temporal and spectral extents, the experimental idler exhibits a global effective chirp rate of approximately 6.6–7.4 THz/ps, about 24–32% below the ideal value of 9.69 THz/ps. Nonlinear propagation, including SPM/XPM, may contribute to this discrepancy, although the strongly non-Gaussian structure introduced by the π phase step prevents attributing it uniquely to SPM.

## Sinusoidal phase transfer

We next probe a distributed phase profile by applying a low-frequency sinusoidal modulation to the signal. The modulation spans approximately 0.6 cycles across the spectral FWHM with an amplitude of 0.56π. This relatively low modulation frequency was chosen deliberately: higher-frequency spectral modulation produces a longer structured temporal waveform and therefore requires a longer pump-overlap window for comparable temporal coverage. Extending this window by further stretching the pump would reduce its peak intensity and consequently the UV conversion efficiency.

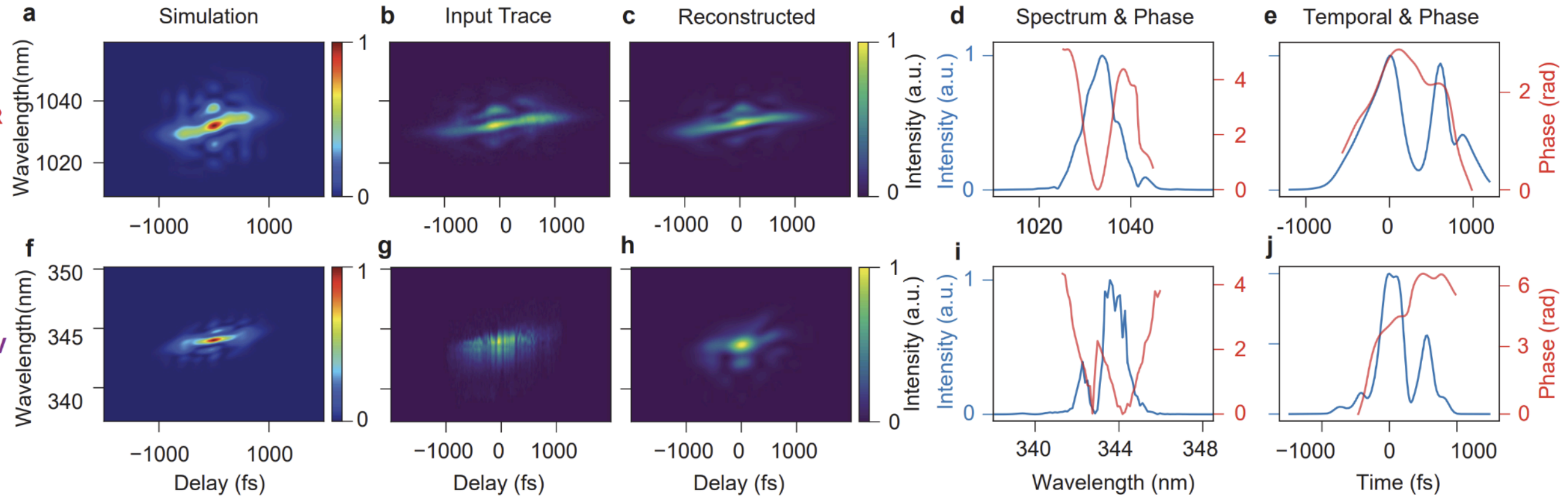

**Fig. 4 | Nontrivial spectral phase transfer with a sinusoidal modulation. a–e,** Simulated and experimental characterization of the infrared signal pulse with a sinusoidal spectral phase. **a,** Simulated TG-FROG trace. **b,c,** Measured and reconstructed TG-FROG traces. **d,** Retrieved spectral amplitude and phase, showing a 0.6-cycle sinusoidal modulation across the spectral FWHM with an amplitude of 0.56π. **e,** Time-domain intensity and instantaneous phase. **f–j,** Corresponding characterization of the ultraviolet idler generated in the gas-filled HCF. **f,** Simulated TG-FROG trace. **g,h,** Measured and reconstructed TG-FROG traces. **i,** Retrieved spectral amplitude and phase of the idler, showing transfer of the sinusoidal spectral phase together with a phase jump at the spectral-intensity minimum. **j,** Time-domain intensity and phase of the idler, showing a corresponding two-pulse temporal structure.

The retrieved signal and idler phases show that the distributed modulation survives frequency conversion (**Fig. 4d,i**). The shaped signal separates into two dominant temporal sub-pulses, with a corresponding two-pulse structure in the UV idler (**Fig. 4e,j**). Unlike the SOD benchmark, this phase profile cannot be represented by a single dispersion coefficient and therefore provides a complementary test of distributed phase transfer.

The idler also develops a spectral minimum accompanied by a rapid phase variation (**Fig. 4i**), consistent with phase-to-amplitude coupling in the pump-conditioned broadband mixing process. We do not further decompose this feature in the phase-transduction analysis below.

## Calibrated phase-conjugating transduction

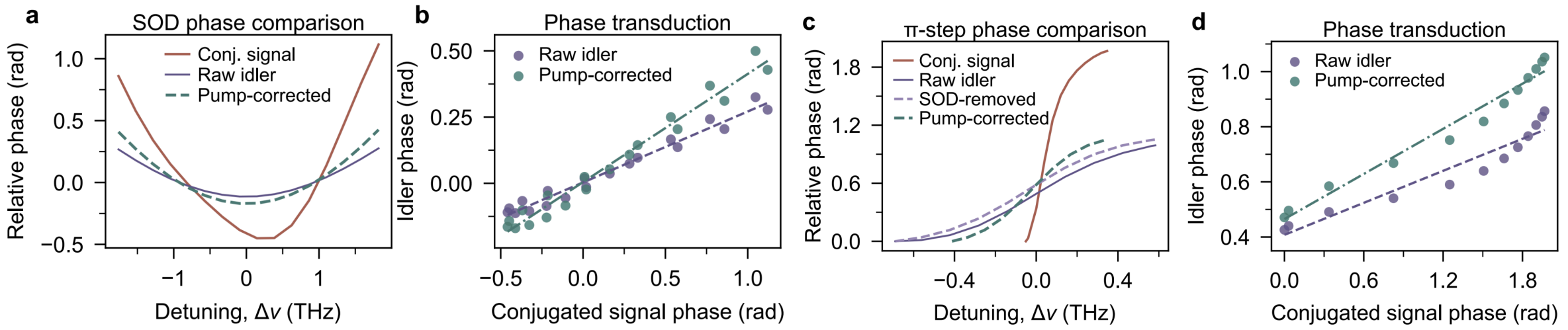

**Fig. 5 | Calibrated quasi-linear phase-transfer response. a,b**, Frequency-matched phase comparison and point-by-point correlation for the SOD case. **c,d**, Corresponding comparison for the localized phase-step case. The raw idler, background-removed or pump-corrected idler, and frequency-matched conjugated signal are shown as appropriate. Dashed lines represent linear fits.

For the SOD and localized-step cases, we next test whether the reconstructed fields obey the same calibrated phase-transduction relation. We map the signal and idler phases onto a common optical-frequency-detuning coordinate according to the frequency-inverted DFWM trajectory. The conjugated signal phase is then compared with the measured idler phase after removal of the model-predicted pump-induced background (**Methods**). In the localized-kernel limit, the two should approach a one-to-one relation.

For the SOD case, the raw idler phase has substantially lower curvature than the conjugated signal because of the deterministic quadratic contribution from the chirped pump (**Fig. 5a**). After removal of the predicted pump-induced background, the corrected idler phase more closely follows the frequency-matched conjugated signal, approaching the one-to-one relation expected in the localized-kernel limit. The reduced curvature of the raw idler therefore reflects the modeled pump-induced phase background together with departures from the ideal point-to-point mapping associated with the finite effective mixing response and propagation-induced nonlinear phase, including SPM.

The localized step probes a higher spectral-modulation-frequency component of the transfer response. After removal of the smooth background, the idler retains the frequency-inverted direction and monotonic evolution of the signal step, but with a broader transition and reduced phase contrast (Fig. 5c). A numerically predicted pump-phase correction applied in post-processing brings the fitted phase-transfer gain closer to unity, while the step remains broadened and reduced in contrast.

This residual discrepancy cannot be assigned uniquely to the nonlinear kernel. Relative to the NIR signal, the UV idler is weaker and is measured with lower effective spectral resolution and signal-to-noise ratio, making the retrieved phase less reliable near abrupt transitions and in low-intensity spectral regions. The measured step response therefore contains contributions from both the physical transfer bandwidth and the diagnostic transfer function. Within these limitations, the frequency-matched phase samples retain an approximately linear correlation (Fig. 5d). The two additional step positions show the same monotonic phase-conjugating response, with position-dependent fitted gain (Supplementary Fig. S7).

### Pump-chirp dependence of phase transfer

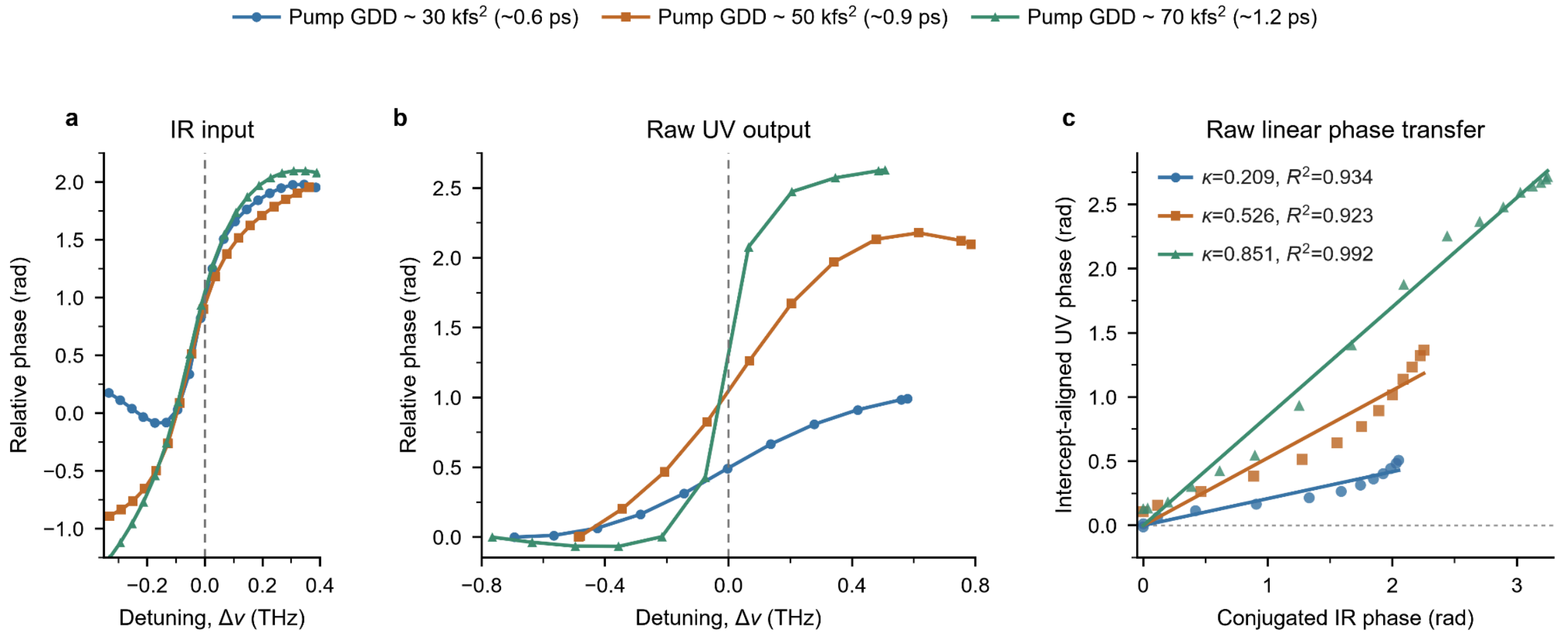

**Fig. 6 | Pump-chirp dependence of localized π-step phase transfer. a–c**, Experimental characterization under three pump-GDD conditions. **a**, Retrieved NIR input phase steps. **b**, Corresponding localized phase components retrieved from the generated UV idler, showing sharper phase transitions at higher pump GDD. **c**, Point-by-point linear fits between the conjugated signal phase and the idler phase, with the fitted slopes quantifying the phase-transfer gain.

To examine the predicted pump-chirp dependence, we measured localized π-step transfer under three pump-GDD conditions. The retrieved UV phase transitions became sharper at higher pump GDD (**Fig. 6b**). To account for variations in the retrieved NIR input phase steps (**Fig. 6a**), we quantified the phase-transfer gain through point-by-point linear fits between the conjugated signal and idler phases, yielding gains approaching unity at higher pump GDD (**Fig. 6c**). Both signal and pump pulse energies were increased at higher pump GDD to compensate for reduced UV generation efficiency and maintain sufficient signal for TG-FROG characterization. Thus, these measurements represent different operating conditions rather than an isolated variation of pump GDD. Nevertheless, the observed trend is consistent with the enhanced time–frequency selectivity predicted by the idealized model (Supplementary Note 2 and Fig. S2).

## Discussion

After correction for the modeled pump-induced phase background, the measured signal-to-idler response approaches a frequency-inverted, phase-conjugating relation. Localized phase features, however, remain broadened and reduced in contrast. The transduction response therefore has a finite spectral-modulation-frequency bandwidth, rather than exhibiting ideal point-to-point transfer for arbitrary phase profiles.

The quasi-linear relation applies to the spectral phase and does not imply universal replication of the temporal intensity waveform. The output spectral amplitude remains conditioned by the pump envelope, phase matching and propagation, so different programmed phases can produce distinct temporal reshaping even when their spectral-phase information is preserved. In the present experiment, finite pump-conditioned bandwidth, residual SPM and GVM, and the lower UV spectral resolution and TG-FROG signal-to-noise ratio may all contribute to the measured departures from the ideal localized-kernel limit. The present measurements do not uniquely separate these contributions.

A fidelity–efficiency trade-off also constrains operation. Increasing pump chirp can enhance local time–frequency selectivity but stretches the pump and lowers its peak intensity. Increasing the injected intensity recovers conversion efficiency but eventually introduces SPM and other propagation-induced phase distortions. The present operating point is further constrained by the sensitivity of the UV TG-FROG, which requires microjoule-level idler pulses for reliable reconstruction. Improved UV detection would allow operation at lower nonlinear phase accumulation and enable a more direct measurement of the transducer bandwidth, for example using sinusoidal phase probes of increasing spectral modulation frequency.

The present HCF-DFWM experiment illustrates how broad phase-matching acceptance and effective localization of the mixing response can support spectral-phase transduction in a co-propagating broadband interaction. The same operator framework applies in principle to other parametric processes that remain linear or conjugate-linear in the shaped input and provide a localized, invertible frequency mapping. Such processes may transfer spectral phase directly or through phase conjugation, with output frequencies above or below the input band. Extension to the DUV and VUV will depend on wavelength-specific dispersion, absorption, ionization and phase-matching constraints.

We demonstrate quasi-linear spectral-phase transduction from the NIR to the UV through DFWM in an argon-filled hollow-capillary fiber, without spatially separating the interacting frequency components. TG-FROG measurements show the transfer of quadratic dispersion, localized phase steps and low-frequency sinusoidal modulation. After correction for the modeled pump-induced phase background, the quadratic and localized-step measurements exhibit approximately linear phase-conjugating responses, although localized features remain broadened and reduced in contrast. These observations are consistent with broadband phase-matching acceptance and an effectively localized pump-conditioned mixing response, identifying the conditions and limitations of spectral-phase transduction in the present system.

## Methods

### Experimental configuration and pulse shaping.

The experiments were driven by a 1032-nm femtosecond laser system operating at 5 kHz. The fundamental pulse had a spectral full width at half maximum (FWHM) of approximately 7.7 nm and a transform-limited duration of approximately 220 fs. The output was divided into a signal arm and a pump arm.

In the signal arm, the 1032-nm pulse was sent through a programmable 4f pulse shaper with a spatial light modulator (SLM) positioned in the Fourier plane. The SLM was used to impose the spectral-phase functions investigated here, including second-order dispersion, localized phase steps and sinusoidal phase modulation. The optical layout, spectral resolution and detailed specifications of the pulse shaper are provided in Supplementary Note 1.

In the pump arm, a fraction of the 1032-nm fundamental was first temporally stretched and subsequently frequency doubled in a β-barium borate (BBO) crystal to generate the 516-nm pump. Residual fundamental light was removed after second-harmonic generation. The stretcher introduced the positive chirp used to establish the pump time–frequency correlation for the DFWM interaction. The pump and shaped signal were recombined with a dichroic optic and coupled into an argon-filled hollow-capillary fiber (HCF), where DFWM generated an ultraviolet idler centred near 344 nm.

$$2\omega_p - \omega_s = \omega_i \tag{M1}$$

The HCF was operated at an argon pressure of approximately 0.61 bar, corresponding to the calculated phase-matching condition for the participating lowest-order modes. The capillary dispersion model, modal phase-matching condition and coupling geometry are given in Supplementary Note 1. The programmed signal dispersion and pump chirp used for individual measurements are specified with the corresponding experimental results.

### TG-FROG characterization and retrieval.

The complex electric fields of the shaped near-infrared signal and generated ultraviolet idler were independently characterized by transient-grating frequency-resolved optical gating (TG-FROG). The fields reported in the main figures were retrieved primarily with the COPRA algorithm. PCGPA retrievals were performed independently as a cross-check of convergence and of the dominant reconstructed phase features, particularly for measurements containing abrupt phase transitions. Unless otherwise noted, the spectral amplitudes and phases shown in the main text correspond to the COPRA retrievals. Details of the TG-FROG geometry and acquisition procedure are provided in Supplementary Note 1.

For quantitative spectral-phase comparison, the retrieved wavelength axes were converted to optical-frequency coordinates before interpolation or fitting. Phase analysis was restricted to the spectrally reliable regions defined from the measured TG-FROG traces as described below. The retrieved phases were unwrapped only within these windows. Constant and linear spectral-phase terms were removed before signal–idler comparison because they represent, respectively, an arbitrary global phase and a relative time origin and do not affect the higher-order or localized spectral-phase structures considered here.

**Broadband DFWM kernel and localized-kernel approximation.**

Under the fixed-pump and weak-depletion approximation, the generated idler field is conjugate-linear in the shaped signal field. For broadband degenerate four-wave mixing, the idler spectral field can be written as

$$\tilde{E}_i(\Omega) = C \iint \tilde{E}_p(\omega_1)\tilde{E}_p(\omega_2)\tilde{E}_s^*(\omega_1+\omega_2-\Omega)\eta(\omega_1,\omega_2,\omega_1+\omega_2-\Omega;\Omega)\, d\omega_1\, d\omega_2 \tag{M2}$$

where $\Omega$ denotes the idler angular frequency, $\omega_1$ and $\omega_2$ are the two pump angular frequencies, and C contains frequency-independent coupling and mode-overlap factors. The function $\eta$ represents the complex phase-matching and propagation response of the interaction. For a uniform interaction in the undepleted limit, the magnitude of its phase-matching contribution contains the usual acceptance factor

$$|\eta| \propto \left|\operatorname{sinc}\left(\frac{\Delta k L}{2}\right)\right| \tag{M3}$$

where L is the interaction length and $\Delta k$ is the frequency-dependent wavevector mismatch. The full capillary dispersion model and calculation of $\Delta k$ are given in Supplementary Note 1.

Equation (M2) can be rewritten exactly as a one-dimensional integral operator acting on the conjugated signal field,

$$\tilde{E}_i(\Omega) = \int H(\Omega,\omega)\tilde{E}_s^*(\omega)\, d\omega \tag{M4}$$

where the two-frequency mixing kernel is

$$H(\Omega,\omega) = C \iint \tilde{E}_p(\omega_1)\tilde{E}_p(\omega_2)\eta(\omega_1,\omega_2,\omega;\Omega)\delta(\Omega+\omega-\omega_1-\omega_2)\, d\omega_1\, d\omega_2 \tag{M5}$$

Using the energy-conservation delta function to perform one of the pump-frequency integrations gives

$$H(\Omega,\omega) = C \int \tilde{E}_p(\omega_1)\tilde{E}_p(\Omega+\omega-\omega_1)\eta(\omega_1,\Omega+\omega-\omega_1,\omega;\Omega)\, d\omega_1 \tag{M6}$$

Under the broad and slowly varying phase-matching approximation, the kernel reduces to a pump-conditioned spectral convolution. The effective local-sampling approximation introduced in the main text is analyzed for chirped Gaussian pump and signal fields in Supplementary Note 2.

## Supplementary information

### Supplementary Note 1: Experimental details

**A. Pulse preparation and shaping.**

The experiments were performed using a 1032-nm femtosecond laser system operating at a repetition rate of 5 kHz, with a maximum average output power of 20 W. The laser was operated below its maximum output power to limit excessive nonlinear effects. The output was split into pump and signal arms, with pulse energies independently adjusted using half-wave plates and beam splitters according to the experimental conditions. Both arms contain a pulse stretcher. In the pump arm, a portion of the fundamental pulse was frequency doubled in a β-barium borate (BBO) crystal to generate the 516-nm pump. The residual fundamental light was subsequently removed, and the pump was recombined with the shaped near-infrared (NIR) signal using a dichroic optic before coupling into the hollow-capillary fibre (HCF).

In the signal arm, the 1032-nm pulse was directed through a programmable 4f pulse shaper incorporating a spatial light modulator (SLM; Santec) at the Fourier plane. The water-cooled SLM is designed for high-power operation, with a specified power-handling capacity of up to 200 W. Its active area measures 15.36 × 9.6 $mm^2$ and contains 1920 × 1200 pixels with an 8-µm pitch.

The pulse-shaper geometry is shown in Supplementary Fig. S1. The incident and reflected beams were vertically displaced at the blazed grating, allowing operation near the Littrow configuration while maintaining spatial separation between the two beams. The NIR pulse had a transform-limited duration of approximately 220 fs and a spectral full width at half maximum (FWHM) of 7.7 nm, corresponding to a width of approximately 13.08 nm at the $1/e^2$ intensity level. Spectral dispersion was provided by a blazed grating with 1,200 grooves $mm^{-1}$ and a cylindrical lens with a focal length of 500 mm, which focused the dispersed spectrum onto the SLM.

The dispersed spectrum occupied approximately 9.9 mm horizontally and 9.5 mm vertically at the Fourier plane. The estimated focal spot diameter on the SLM is

$$D_{\mathrm{SLM}} = \frac{4M^2 f\lambda}{\pi D_{\mathrm{in}}} \approx 90\,\mu\mathrm{m} \tag{S1.1}$$

which exceeds the SLM pixel pitch of 8 µm. The number of independently resolvable spectral elements is therefore estimated to be approximately $\frac{9.9\,\mathrm{mm}}{90\,\mu\mathrm{m}} \approx 110$.

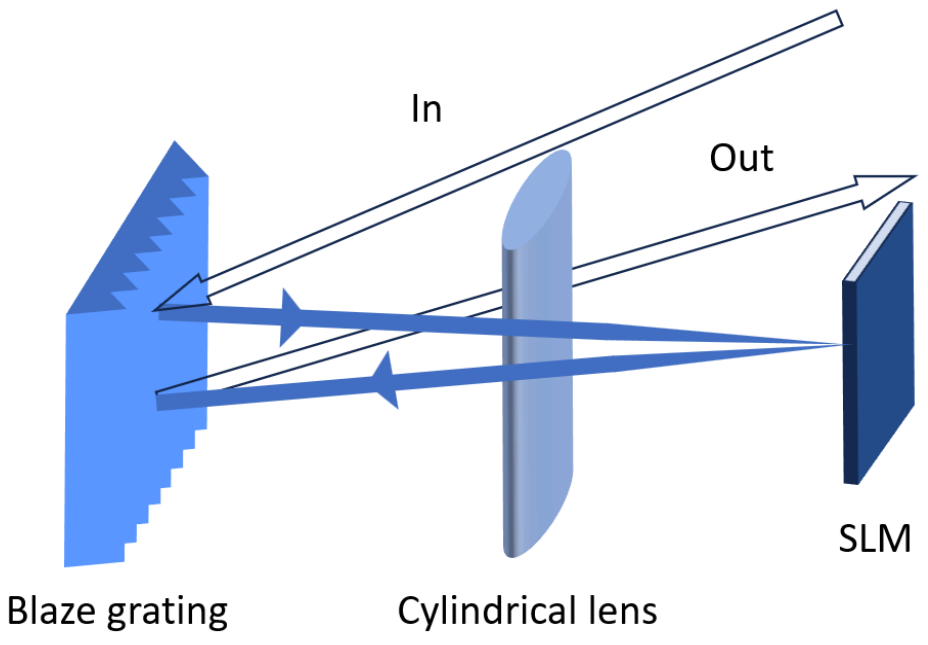

**Fig. S1 | Schematic of the NIR pulse shaper.**

**B. Gas-filled HCF coupling, frequency mixing phase matching.**

Beam-expanding telescopes were used in both the pump and signal arms before coupling into the HCF. The two beams shared a final focusing lens with a focal length of 500 mm. The coupling was optimized for the visible pump, resulting in measured transmission values of 22% for the pump and 0.6% for the signal. The HCF was 1 m long, with an inner-core diameter of 100 µm.

Under the undepleted-pump approximation, the magnitude of the phase-matching acceptance function is given by

$$|\eta(\omega_s, \omega_p, \omega_i)|^2 = \mathrm{sinc}^2 \left[\frac{1}{2}\Delta k(\omega_s, \omega_p, \omega_i) L\right] \tag{S1.2}$$

where $\Delta k$ is the wavevector mismatch, and $L$ represents the HCF length. $\omega_s, \omega_p, \omega_i$ represents the signal, pump, and idler frequency, respectively. The wavevector mismatch $\Delta k$ can be expressed as

$$\Delta k = k_s + k_i - 2k_p = \Delta k_{\mathrm{material}} - \Delta k_{\mathrm{mode}} = \frac{2\pi p}{\lambda_s}(\delta_s + 3\delta_i - 4\delta_p) - \frac{\lambda_s}{4\pi a^2}\left(u_s^2 + \frac{u_i^2}{3} - u_p^2\right) \tag{S1.3}$$

where $p$, $a$, $u_n$ are the gas pressure, core radius, and the modal constant, respectively. The gas refractive index is expressed as $n_{s,i,p} = 1 + p\delta_{s,i,p}$, where the refractive-index contribution depends on gas pressure $p$. The material-dispersion contribution to the phase mismatch scales with pressure, whereas the waveguide contribution is determined by the capillary geometry and mode. Phase matching is achieved when these contributions balance, such that Δk = 0. For the signal and pump propagating in the lowest-order $EH_{11}$ mode, the calculated and experimental phase-matching pressure is 0.61 bar.

### C. TG-FROG characterization

The generated UV beam was divided into three paths using a three-hole mask. One beam passed through a delay stage before all three beams were focused onto a nonlinear crystal to generate the transient-grating (TG) signal, which was recorded using a spectrometer.

The measured TG-FROG traces were retrieved using both the principal components generalized projections algorithm (PCGPA) and the COPRA algorithm. Retrievals were performed with different initial conditions to assess convergence and the consistency of the reconstructed traces and spectra. For measurements containing localized π-phase steps, convergence was particularly sensitive to the initial conditions near abrupt phase transitions. Multiple retrievals were therefore performed until the reconstructed traces and dominant spectral features were consistent across the algorithms.

## Supplementary Note 2: Kernel localization in broadband DFWM

Under the fixed-pump and weak-depletion approximations, the generated idler field is conjugate-linear in the input signal field. As described by Eqs. (M4)–(M6), the transformation is

$$\tilde{E}_i(\Omega) = \int H(\Omega, \omega)\tilde{E}_s^*(\omega)\, d\omega \tag{S2.1}$$

Here, $\omega$ and $\Omega$ are the signal and idler angular frequencies, respectively, consistent with the main text. When phase matching is broad and slowly varying over the interacting bandwidths, the mixing kernel reduces to the spectral self-convolution of the pump,

$$H(\Omega, \omega) \simeq CQ(\xi), \qquad \xi = \Omega + \omega - 2\omega_{p0} \tag{S2.2}$$

Here $\omega_{p0}$ is the pump carrier angular frequency, $Q$ is the self-convolution of the pump spectrum expressed in angular-frequency detuning, and $Q$ is proportional to the Fourier transform of $q(t) = E_p^2(t)$, with $E_p(t)$ denoting the pump envelope. Constant Fourier-convention factors are absorbed into $C$. We use carrier frequencies satisfying $\Omega_{i0} + \omega_{s0} = 2\omega_{p0}$.

The ideal point-to-point spectral mapping corresponds to a localized sampling kernel,

$$H(\Omega, \omega) \simeq G(\Omega)\delta\left[\omega - (2\omega_{p0} - \Omega)\right] \tag{S2.3}$$

which yields $\tilde{E}_i(\Omega) \simeq G(\Omega)\tilde{E}_s^*(2\omega_{p0} - \Omega)$. The delta function describes the ideal selectivity of the mixing kernel with respect to input frequency at fixed output frequency. Such an effective local-sampling response can arise either from spectral-amplitude confinement or from phase-dependent coherent interference.

### A. Amplitude-induced localization: narrowband pump

The most direct route to kernel localization is to narrow the pump spectrum. Under broad phase matching, $H(\Omega, \omega)$ is proportional to $Q(\xi)$. As the pump bandwidth decreases, $|Q(\xi)|$ becomes concentrated near $\xi = 0$. In the monochromatic-pump limit, an appropriately normalized $Q$ approaches $\delta(\xi)$, producing the frequency-inverted mapping described by Eq. (S2.3). Here the local sampling originates from the narrowing of the spectral magnitude itself.

### B. Phase-induced effective localization: dispersive pump

To illustrate a phase-induced mechanism, consider a Gaussian pump with fixed spectral intensity and quadratic spectral phase. Let $\sigma_p$ denote its spectral-intensity RMS width in angular-frequency units and $\beta_p$ its GDD. The effective two-pump spectrum remains Gaussian, with $\sigma_q = \sqrt{2}\sigma_p$ and $\beta_q = \beta_p/2$. The complex kernel is then

$$H(\Omega, \omega) = B_{\beta_p} \exp\left(-\frac{\xi^2}{4\sigma_q^2}\right) \exp\left(i\frac{\beta_p \xi^2}{4}\right) \tag{S2.4}$$

where $\xi = \Omega + \omega - 2\omega_{p0}$, and $B_{\beta_p}$ contains the chirp-dependent overall amplitude. Increasing pump GDD leaves the normalized magnitude of this Gaussian kernel unchanged but increases its quadratic spectral-phase curvature.

The effect of this phase on the *integral action* of the kernel can be seen from stationary phase. For fixed $\Omega$, changing the integration variable to $\xi$ gives an oscillatory factor $\exp(i\beta_p \xi^2/4)$ with a single stationary point at $\xi = 0$. In the limit $|\beta_p| \to \infty$, contributions away from this point cancel. Provided that the signal spectrum remains fixed and sufficiently smooth over the stationary-phase region, the leading-order result is

$$\tilde{E}_i(\Omega) \sim B_{\beta_p}\sqrt{\frac{4\pi}{|\beta_p|}} \exp\left[i \ \mathrm{sgn}(\beta_p)\frac{\pi}{4}\right] \tilde{E}_s^*(2\omega_{p0} - \Omega) = C(\beta)\tilde{E}_s^*(2\omega_{p0} - \Omega) \tag{S2.5}$$

The corresponding idler phase is,

$$\phi_i(\Omega) \simeq -\phi_s(2\omega_{p0} - \Omega) + \phi_C \tag{S2.6}$$

where $\phi_C$ is the phase of the effective complex prefactor. The convolution therefore approaches a frequency-inverted sampling operation without a corresponding narrowing of the kernel magnitude. This result is asymptotic and does not imply exact transfer of abrupt phase features at finite pump GDD.

The same behavior can be examined in the time–frequency domain. For the Gaussian effective pump field $q(t) = E_p^2(t)$, its Wigner distribution is

$$W_q(t, \xi) \propto \exp\left(-\frac{\xi^2}{2\sigma_q^2}\right) \exp\left[-\frac{(t - \beta_p \xi/2)^2}{2\tau_q^2}\right] \tag{S2.7}$$

where $\tau_q = 1/(2\sigma_q)$ is the transform-limited temporal-intensity RMS width of $q(t)$. The tilted ridge $t = \beta_p \xi/2$ represents the pump-induced time–frequency correlation. At any fixed time, the local frequency distribution has the same RMS width,

$$\sigma_{\mathrm{local},\omega} = \left[\frac{1}{\sigma_q^2} + \frac{\beta_p^2}{4\tau_q^2}\right]^{-1/2} \quad \text{(S2.8)}$$

This width decreases with increasing $|\beta_p|$, although the time-integrated spectral magnitude retains its original normalized width. A fixed-time slice therefore visualizes the increasing local pump-frequency selectivity rather than a narrowing of the global spectrum.

To illustrate the effect within a finite pump–signal overlap, Fig. S2 plots a conditional time–frequency mixing response in the signal–idler frequency plane. Under the instantaneous DFWM relation $E_i(t) = Cq(t)E_s^*(t)$, the overlap can be expressed, up to a normalization constant, as

$$M(\Omega, \omega) \propto \int W_q(t, \xi) W_s(t, \omega - \omega_{s0})\, dt \quad \text{(S2.9)}$$

where $W_s$ is the Wigner distribution of the reference signal envelope. Integration of $M$ over the signal angular-frequency coordinate recovers the idler spectral intensity, up to the same Fourier normalization, in this instantaneous model. For the Gaussian fields considered here, this conditional distribution is nonnegative and provides an intuitive visualization of the effective mixing pathways.

For Fig. S2, the pump and reference signal are Gaussian pulses with transform-limited intensity FWHM durations of 160 fs and 220 fs, respectively. The pump spectrum and phase-matching response are held fixed, the signal remains transform limited, and only the pump GDD is varied. With $\sigma_s$ the signal spectral-intensity RMS width in angular-frequency units and $\tau_s = 1/(2\sigma_s)$, the signal-axis FWHM at the central idler frequency, expressed in ordinary frequency (Hz), is

$$\Delta\nu_{\mathrm{cond}} = \frac{\sqrt{2\ln 2}}{\pi}\sigma_{\mathrm{cond},\omega}$$
$$\sigma_{\mathrm{cond},\omega}^{-2} = \sigma_s^{-2} + \sigma_q^{-2} + \frac{\beta_p^2}{4(\tau_s^2 + \tau_q^2)} \quad \text{(S2.10)}$$

The axes in Fig. S2 use ordinary-frequency detunings $\Delta\nu_s = \frac{\omega - \omega_{s0}}{2\pi}$ and $\Delta\nu_i = \frac{\Omega - \Omega_{i0}}{2\pi}$, reported in THz. At $\Delta\nu_i = 0$, the calculated conditional FWHM decreases from 1.784 THz at zero pump GDD to 1.002 THz at 65,000 $fs^2$ and 0.720 THz at 100,000 $fs^2$. The individually normalized maps show the increasing concentration of the conditional overlap around $\Delta\nu_i = -\Delta\nu_s$.

Fig. S2 is intended as a visualization of pump-induced time–frequency selectivity, not as an independent definition of the universal mixing kernel or of phase-transfer resolution. The conditional response $M$ depends on the chosen reference signal, whereas $H$ is determined by the pump and the phase-matching response. The illustrated trend is consistent with the improved preservation of localized spectral-phase features at larger pump GDD in Fig. 6. The reduced model neglects SPM, XPM, GVM, and pump depletion.

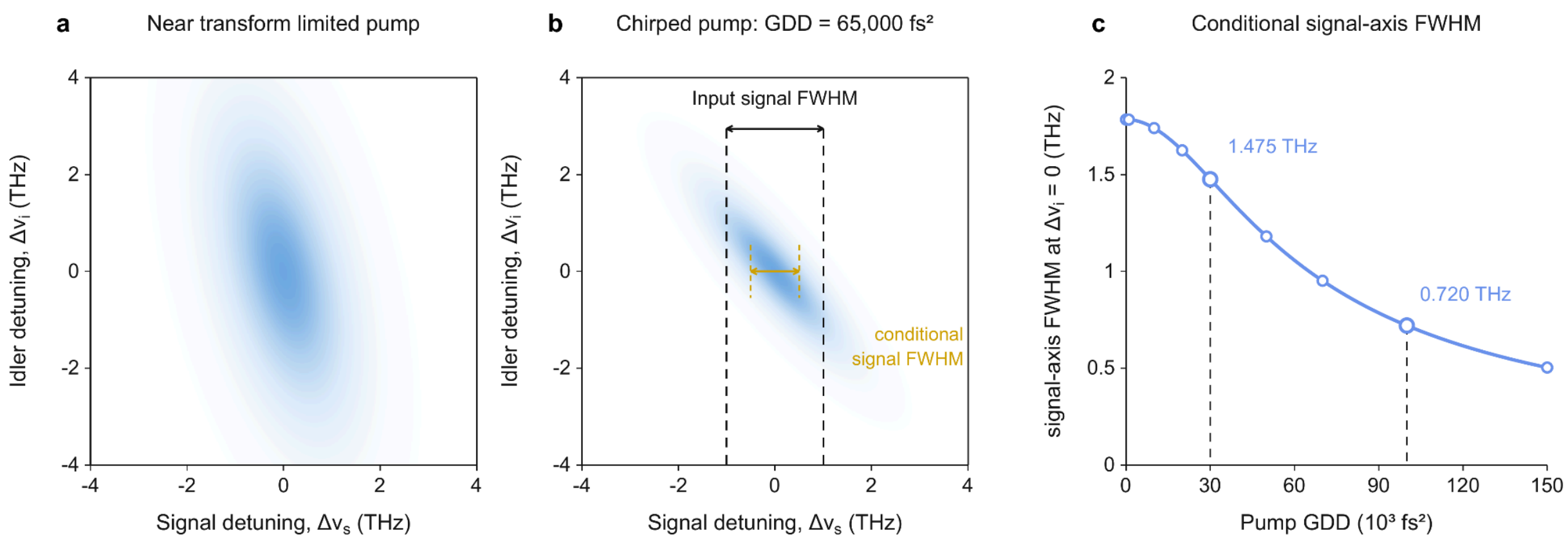

**Fig. S2 | Pump-chirp dependence of the conditional DFWM mixing response. a,b**, Individually normalized time–frequency overlap maps M(Δvi, Δvs), calculated from Eq. (S2.9) for pump GDDs of 0 and 65,000 fs², respectively. The pump spectrum and reference-signal parameters are held fixed. Increasing pump GDD concentrates the overlap near the frequency-inverted trajectory Δvi = −Δvs. The grey dashed lines in b mark the spectral FWHM of the input signal, while the orange markers indicate the conditional signal-axis FWHM at Δvi = 0 (1.002 THz). **c**, Conditional signal-axis FWHM at Δvi = 0 as a function of pump GDD, decreasing from 1.784 THz at zero GDD to 0.720 THz at 100,000 fs².

Figure S2 shows the effect of pump GDD on the reference-signal-weighted mixing response. With the pump spectrum and reference signal held fixed, increasing pump GDD concentrates the conditional overlap near the frequency-inverted trajectory and reduces its signal-axis FWHM. This narrowing reflects the pump time–frequency correlation rather than a change in the spectral magnitude of H. Although the trend agrees qualitatively with the improved phase-step preservation in Fig. 6, the calculated width is not a direct measure of experimental phase-transfer resolution, since the model neglects propagation-induced nonlinear phase and assumes a transform-limited Gaussian reference signal.

## Supplementary Note 3: Additional experimental results for nontrivial π-step transfer

### A. Centre-positioned π-step

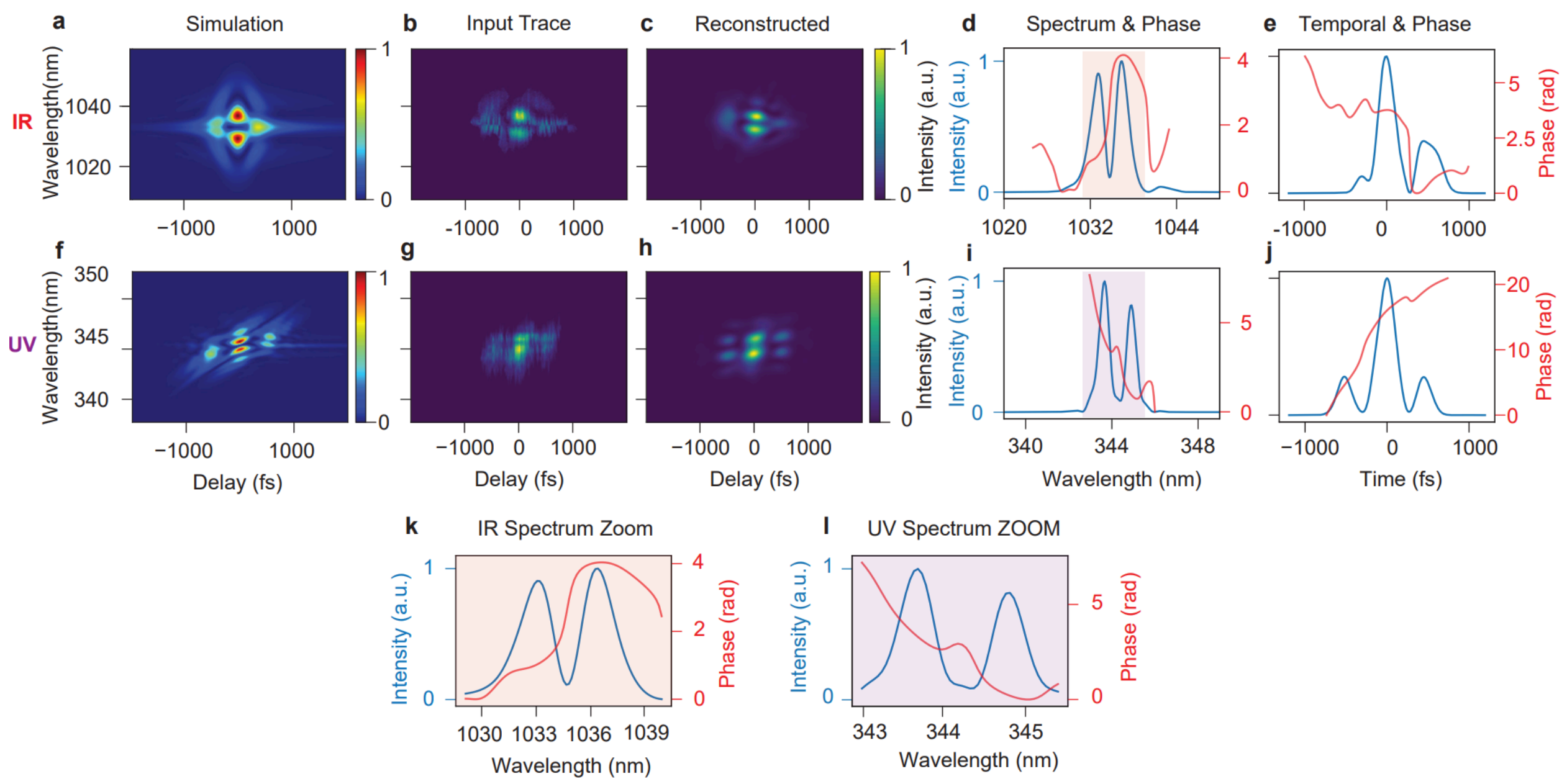

**Fig. S3 | Spectral-phase transfer with a centre-positioned π-step. a–e**, Simulated and experimental characterization of the NIR signal pulse with a π-step superimposed on second-order dispersion (SOD). **a**, Simulated TG-FROG trace. **b,c**, Measured and reconstructed TG-FROG traces. **d**, Retrieved spectral amplitude and phase. **e**, Temporal intensity and instantaneous phase. **f–j**, Corresponding characterization of the generated ultraviolet idler. **f**, Simulated TG-FROG trace. **g,h**, Measured and reconstructed TG-FROG traces. **i**, Retrieved spectral amplitude and phase. **j**,

Temporal intensity and instantaneous phase. **k,l**, Enlarged views of the spectral amplitude and phase in **d** and **i**, showing that the phase step remains near the centre of the spectrum after transfer to the idler.

## B. Right-positioned $\pi$-step

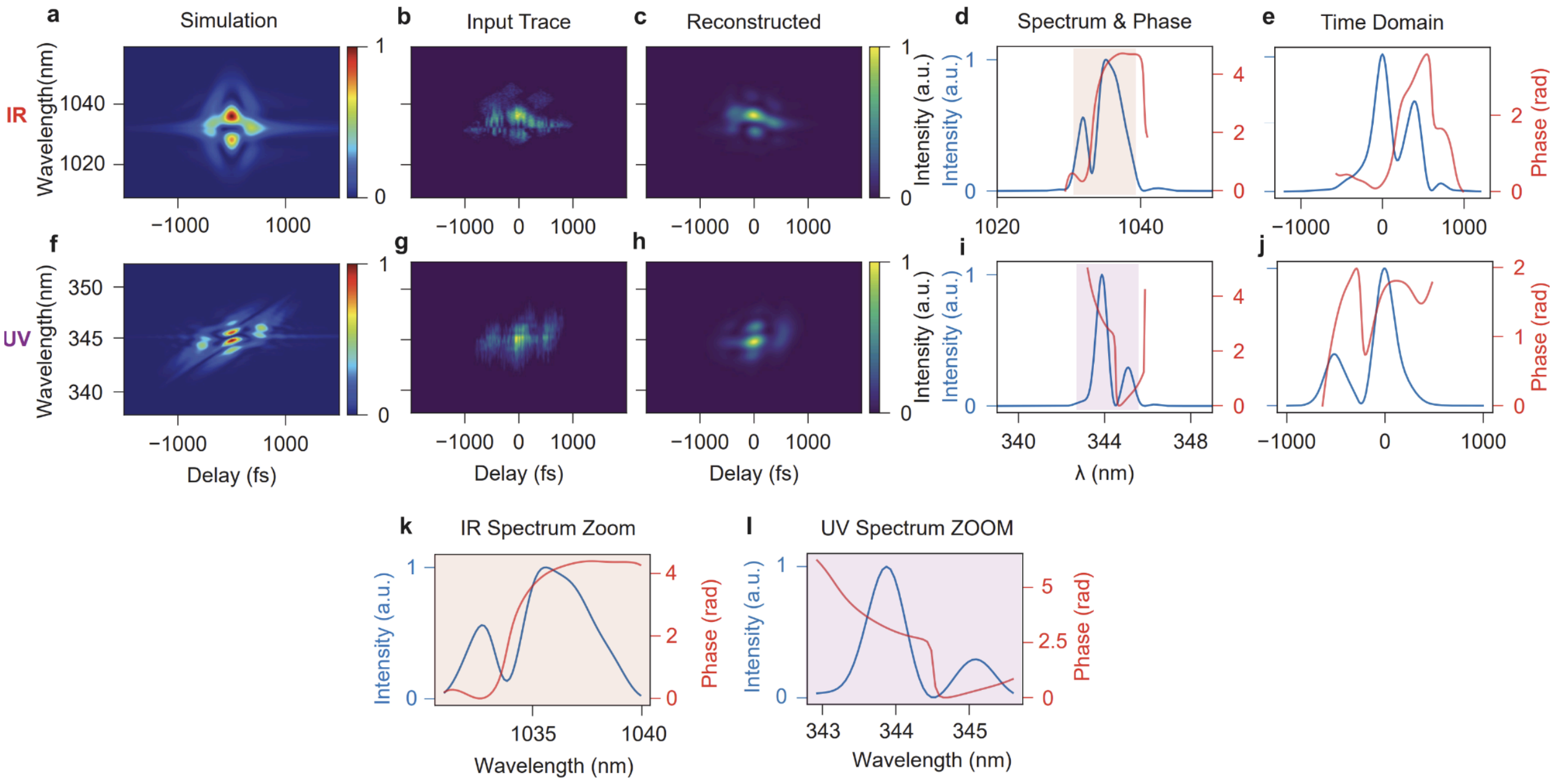

**Fig. S4 | Spectral-phase transfer with an off-centre $\pi$-step. a–e**, Simulated and experimental characterization of the near-infrared signal pulse with a $\pi$-step superimposed on second-order dispersion (SOD). **a**, Simulated TG-FROG trace. **b,c**, Measured and reconstructed TG-FROG traces. **d**, Retrieved spectral amplitude and phase. **e**, Temporal intensity and instantaneous phase. **f–j**, Corresponding characterization of the generated ultraviolet idler. **f**, Simulated TG-FROG trace. **g,h**, Measured and reconstructed TG-FROG traces. **i**, Retrieved spectral amplitude and phase. **j**, Temporal intensity and instantaneous phase. **k,l**, Enlarged views of the spectral amplitude and phase in **d** and **i**, showing that the phase step is transferred from the short-wavelength side of the signal spectrum to the long-wavelength side of the idler spectrum, consistent with the frequency-inverted DFWM interaction.

## C. Comparison of all three step positions

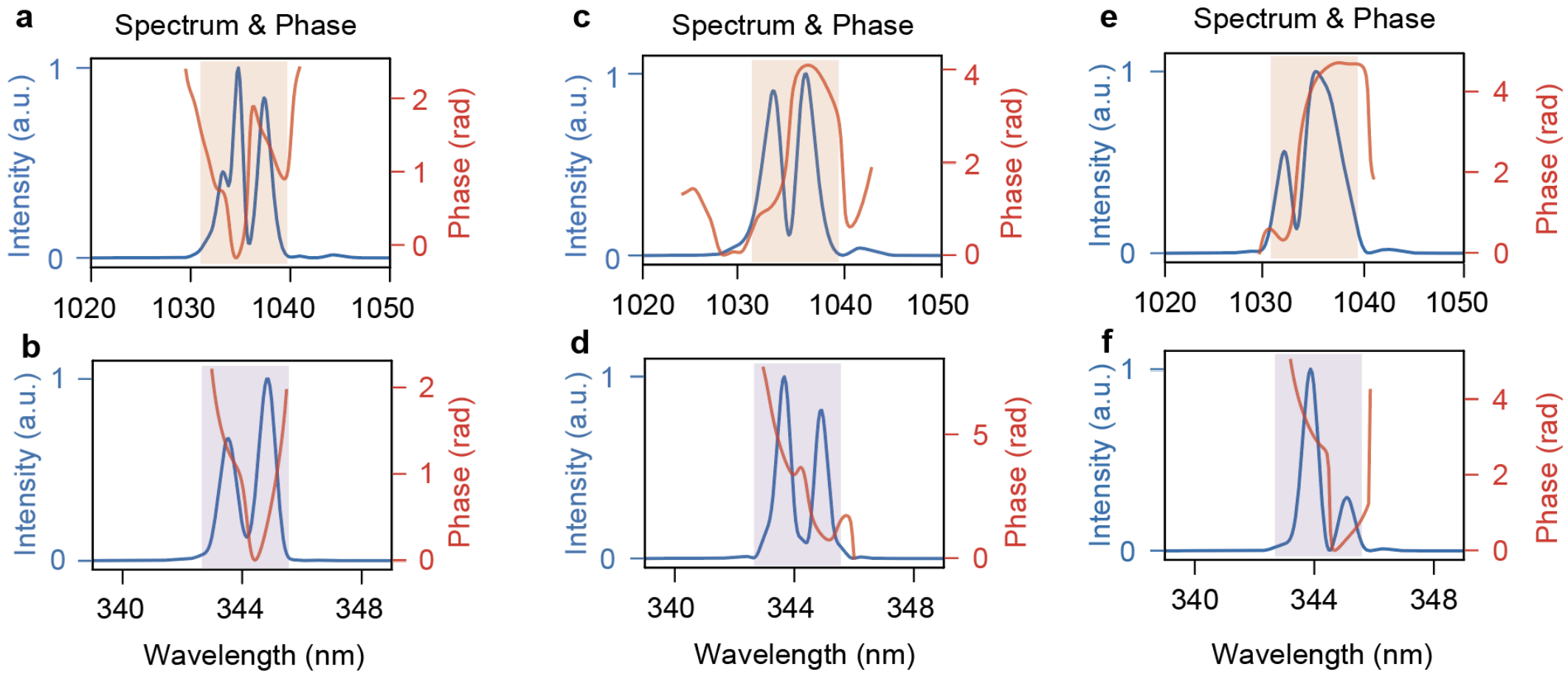

**Fig. S5 | Comparison of spectral-phase transfer for three $\pi$-step positions. a,c,e**, Retrieved spectral amplitudes and phases of the NIR signal with the $\pi$-step applied to the long-wavelength side, centre and short-wavelength side of the spectrum, respectively. **b,d,f**, Corresponding spectral amplitudes and phases of the generated UV idler.

### D. Comparison of conjugated signal and idler phases of all three pi-step positions.

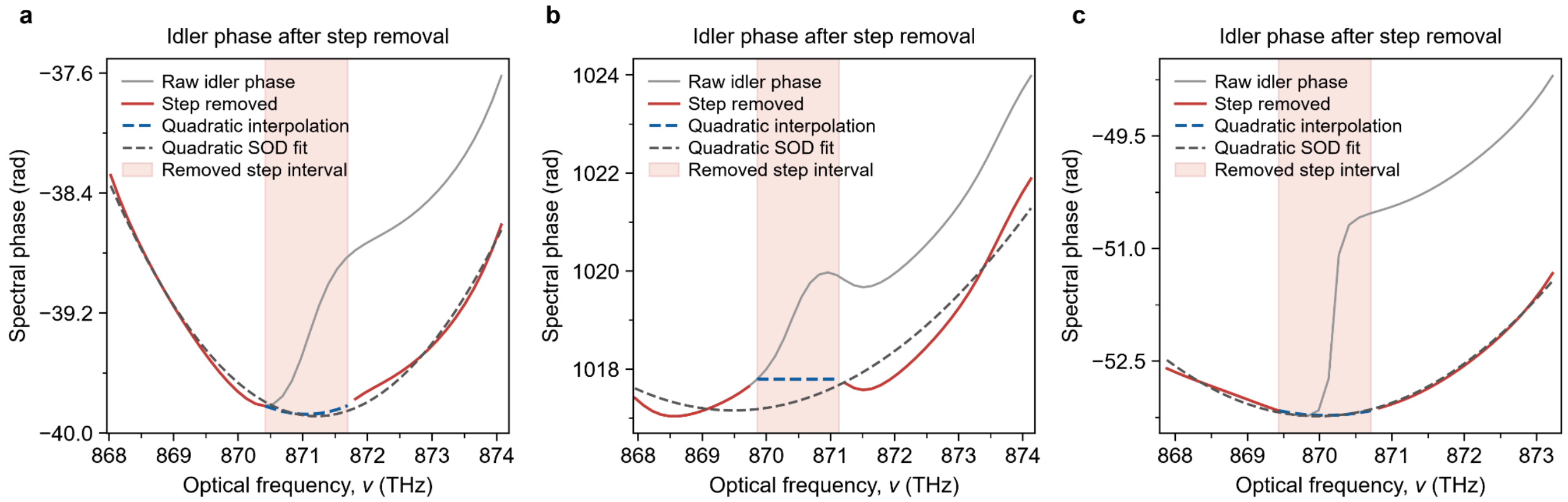

**Fig. S6 | Idler spectral-phase decomposition for three π-step positions. a,b,c**, Spectral-phase decomposition of the generated idler for phase steps positioned on the left, at the centre and on the right of the signal spectrum, respectively.

### E. Position dependence of the localized phase-transfer response

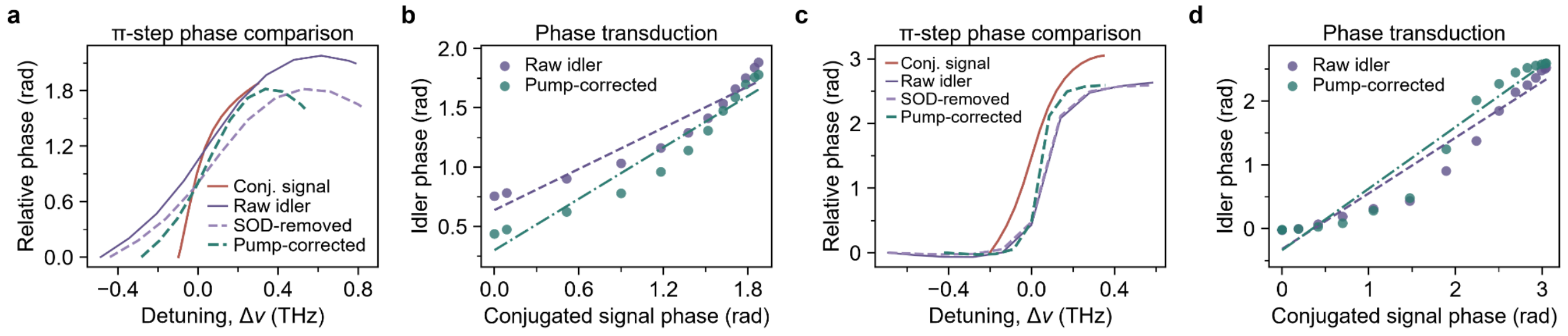

**Fig. S7 | Calibrated quasi-linear phase-transfer response for the middle- and right-localized π-step modulation.** **a,c**, Frequency-matched conjugated signal phase and reconstructed idler phases for the middle, right localized phase step. The raw idler, the idler after removal of its smooth SOD background and the pump-corrected idler are shown for comparison. **b, d**, Point-by-point correlation between the conjugated signal phase and the raw or pump-corrected idler phase for the middle and right step positions. Dashed lines represent linear fits.

To test the robustness of the localized phase transfer, the programmed phase step was placed at two additional positions within the signal spectrum. In both cases, the reconstructed idler retains the frequency-inverted step and exhibits an approximately linear point-by-point correlation with the conjugated signal phase. The fitted phase-transfer gain varies with step position and slightly exceeds unity in one corrected dataset. Because this coefficient is a local phase-response slope rather than an energy gain, values above unity are not intrinsically unphysical. The position dependence may reflect the complex, nonuniform pump-conditioned transfer kernel, including coherent phase–amplitude coupling, as well as differences in spectral weighting and UV retrieval uncertainty. The present measurements do not uniquely separate these contributions.

## Supplementary Note 4: Retrieval, fitting and sources of discrepancy

### A. TG-FROG retrieval and fitting procedure

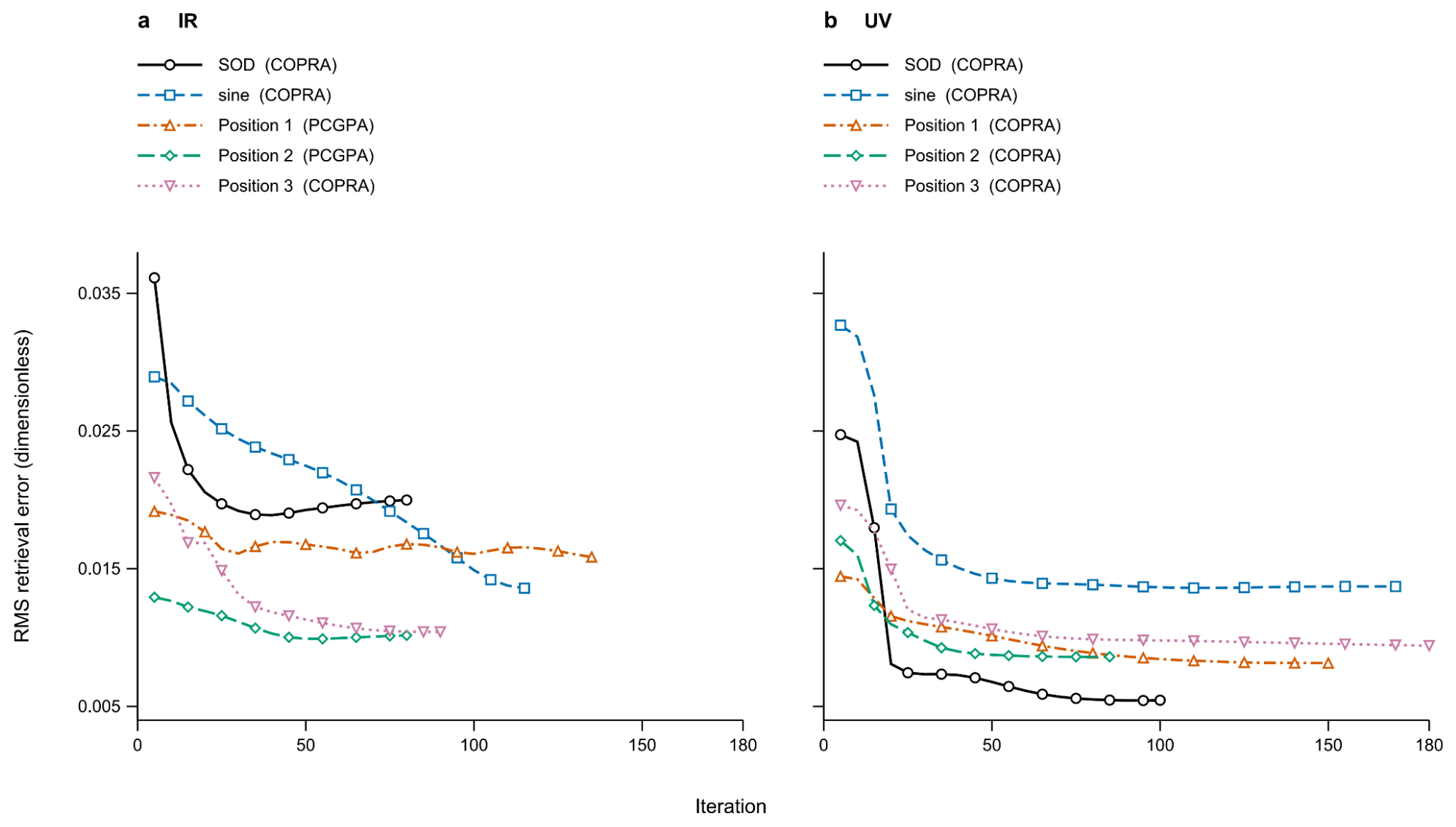

**Fig. S8 |** TG-FROG retrieval convergence for the reported measurements. RMS trace-reconstruction error as a function of iteration for the NIR signal (a) and UV idler (b), including the SOD, sinusoidal and localized-step measurements.

All reported retrievals converge to stable residual levels, with comparable convergence obtained from repeated initial conditions and, where applicable, independent COPRA and PCGPA retrievals. The residual is used here as a convergence diagnostic rather than as a direct uncertainty estimate for the retrieved spectral phase. Abrupt phase transitions and low-intensity UV spectral regions remain more sensitive to finite spectrometer resolution, noise and retrieval initialization.

**B. Frequency mapping, pump-phase calibration and phase-transfer fitting.**

All quantitative signal–idler phase comparisons were performed in optical-frequency rather than wavelength coordinates. The measured wavelength axes were converted according to

$$\nu = \frac{c}{\lambda} \tag{S4.1}$$

The spectral analysis window for each measurement was defined directly from the measured TG-FROG trace. The spectral marginal of the raw trace was obtained by integrating over delay, and the contiguous region for which this marginal exceeded 10% of its peak value was retained. The same window was then applied to the retrieved spectral field. This threshold excludes low-signal spectral regions in which the retrieved phase is weakly constrained and especially sensitive to spectrometer noise and retrieval instability.

Within each retained window, the intensity centroid of the reconstructed spectrum was calculated as

$$\bar{\nu}_j = \frac{\int_{W_j} \nu I_j(\nu)\, d\nu}{\int_{W_j} I_j(\nu)\, d\nu}, \qquad j = s, i \tag{S4.2}$$

and the relative-frequency coordinates were defined by

$$\Delta\nu_s = \nu_s - \bar{\nu}_s, \qquad \Delta\nu_i = \nu_i - \bar{\nu}_i \tag{S4.3}$$

The signal and idler were then frequency matched using the phase-conjugating inversion $\Delta v_s = -\Delta v_i$ . The corresponding mapped signal phase on the idler detuning grid was

$$\phi_{s,\mathrm{map}}(\Delta\nu_i) = -\phi_s(-\Delta\nu_i) \tag{S4.4}$$

Centroid-based detuning alignment was used instead of enforcing the mapping from the absolute spectrometer wavelength calibration. In the present measurements, the NIR signal and UV idler lie close to opposite edges of the spectral ranges used by their TG-FROG spectrometers, where a small absolute wavelength offset can produce an appreciable error in the inferred optical-frequency separation. Referencing each reconstructed spectrum to its measured centroid removes sensitivity to a constant calibration offset while preserving the relative frequency inversion that is relevant to the phase-transfer comparison.

The reconstructed idler phase contains both the conjugated signal contribution and a smooth quadratic background originating from the chirped pump. This pump-induced contribution was estimated with the forward-convolution model rather than by subtracting the full nominal pump GDD directly. For each measurement, the temporal-overlap fraction between the reconstructed shaped signal and the stretched pump was first evaluated. The forward convolution was then used to calculate the idler phase generated by the fraction of the chirped pump sampled over that effective interaction window. The quadratic component of this modeled idler background defines the effective pump-to-idler chirp, $\varphi_{□\rightarrow i}^{(2)}$, for that measurement. The pump-corrected idler phase was obtained by subtracting only the corresponding quadratic phase,

$$\phi_{i,\mathrm{corr}}(\Delta\nu) = \phi_i(\Delta\nu) - \frac{1}{2}\phi_{p\rightarrow i}^{(2)}(2\pi\Delta\nu)^2 \tag{S4.5}$$

The correction is determined from the reconstructed signal and pump parameters and is not adjusted to maximize agreement between the signal and idler phases. It therefore removes only the modeled quadratic pump background; any remaining attenuation, broadening or nonlinear distortion is retained in the comparison. This correction removes the deterministic chirp associated with the temporally sampled pump but does not alter the measured idler spectral amplitude and does not attempt to restore localized phase components that may have been smoothed or converted into amplitude modulation by the finite mixing kernel.

For the localized phase-step measurements, the smooth dispersive background was determined independently of the step transition. The spectral interval containing the transition was excluded, the remaining phase was fitted with a quadratic function, and this fit was extrapolated across the excluded interval. Subtracting the extrapolated background isolated the localized phase component used to determine the step amplitude and transition position. The same procedure was applied to all programmed step positions.

Phase-transfer gains were extracted from frequency-matched phase pairs within the common reliable signal–idler window, defined by the overlap of the two 10%-threshold masks after frequency mapping. For each retained frequency sample,

$$x_j = \phi_{s,\mathrm{map}}(\Delta\nu_j), \qquad y_j = \phi_i(\Delta\nu_j) \text{ or } \phi_{i,\mathrm{corr}}(\Delta\nu_j) \tag{S4.6}$$

and an ordinary least-squares fit was performed using

$$y_j = g x_j + b \tag{S4.7}$$

where g is the phase-transfer gain and b accounts for any residual constant phase offset. An ideal calibrated phase-conjugating response corresponds to g = 1. Because g quantifies the slope of a spectral-phase response rather than an optical power or energy ratio, values slightly above unity are not intrinsically unphysical. The reported deviations from unity should be interpreted together with the finite transfer bandwidth and the larger spectral and retrieval uncertainty of the weaker UV idler. Here, $g$ quantifies the transferred phase contrast; quasi-linearity refers to the approximately point-to-point phase relation itself and does not require an exactly unity-valued gain.